\documentclass[runningheads]{llncs}
\usepackage[T1]{fontenc}
\usepackage{amsmath}
\usepackage{amssymb}
\usepackage{enumitem}
\usepackage{hyperref}
\usepackage{mathtools}
\usepackage{geometry}
\usepackage{graphicx}
\begin{document}
\title{Towards LLM-based Generation of Human-Readable Proofs in Polynomial Formal Verification\\ {\small (Invited Paper)}}
%
%
\author{Rolf Drechsler\orcidID{0000-0002-9872-1740}
}

\authorrunning{R. Drechsler}
\titlerunning{Towards LLM-based Generation of Human-Readable Proofs in PFV}
%
\institute{University of Bremen/DFKI, Bremen, Germany \\   
\email{drechsler@uni-bremen.de} \\
\url{http://www.rolfdrechsler.de} 
}
%

\maketitle              
\begin{abstract}
Verification is one of the central tasks in circuit and system design. While simulation and emulation are widely used, complete correctness can only be ensured based on formal proof techniques. But these approaches often have very high run time and memory requirements. Recently, {\em Polynomial Formal Verification} (PFV) has been introduced showing that for many instances of practical relevance upper bounds on needed resources can be given. But proofs have to be provided that are human-readable.

Here, we study how modern approaches from {\em Artificial Intelligence} (AI) based on {\em Large Language Models} (LLMs) can be used to generate proofs that later on can be validated based on reasoning engines. Examples are given that show how LLMs can interact with proof engines, and directions for future work are outlined.  
\keywords{Polynomial Formal Verification \and LLM \and Proof Generation.}
\end{abstract}
\section{Introduction}

With the ever-increasing complexity of modern circuits and systems, ensuring the functional correctness is a highly relevant task. Classical approaches based on simulation and emulation are not sufficient due to their limited resulting coverage in the verification task. Only formal proof techniques can guarantee that the systems operate 100\% correct. But the underlying methods, like BDDs, SAT or SMT, are known to be computationally very expensive (see e.g.~\cite{Springer2004,Springer2017}). 

A possible solution to this dilemma is {\em Polynomial Formal Verification} (PFV) \cite{DDECS2021,ICCAD2022}: For specific classes of circuits, like adders and multipliers, implementations are identified together with a corresponding proof engine such that efficient upper bounds on run time and memory requirement can be ensured. This requires first a theoretical analysis and a formal proof followed by an efficient implementation. By this, fast adder designs with more than one thousand bits could be completely verified based on BDDs in a few CPU seconds. 

While PFV itself is very powerful, so far, its lacks automation of the proof process. Recently in \cite{MEMOCODE2023,PhilTrans2025} approaches have been proposed for the automatic generation of human-readable proofs. These are based on a search on the underlying graphs of BDDs. 

In this paper, we propose to enhance the proof generation process by techniques from {\em Artificial Intelligence} (AI), namely {\em Large Language Models} (LLMs). These have been studied in various applications in circuit design and have also recently been investigated in the context of verification \cite{ATS2024}. We show some examples for generation of human-readable proofs. Due to the power of today's LLMs -- in the following ChatGPT 4o is used -- easily understandable text can be created. But due to the hallucinations in the AI tool, finally the correctness has to be proven. We show some examples to discuss the consequences for future work. 

\section{Examples}

One of the simplest but also in this context often used proof technique is {\em induction}. This is also the technique applied in \cite{MEMOCODE2023,PhilTrans2025} for BDDs. 

In the following, it is briefly described how LLMs can be used, while the detailed questions and answers are reported in Section \ref{se:appendix} {\em Appendix}. The corresponding question/prompt is given by {\bf Qx:} followed by the answer {\bf A:}.

\begin{description}

\item[Induction proof:] Not surprisingly, the basic of mathematical induction can be easily summarized (see {\bf Q1}), where the induction step is given from $n$ to $n+1$. 

\item[\boldmath\( f_2 = x_1x_2 + x_3x_4 + \cdots + x_{2n-1}x_{2n} \):] In a next step, induction is applied to function $f_2$. Here, it is important to do the induction step not only for a single but for two variables. This is also suggested by the tool (see {\bf Q2}). This can be expected, since this function is very well known and well-studied in the BDD community. Already in the original paper by Bryant, where BDDs have been introduced in its often used ordered and reduced form, this function has been discussed \cite{TC1986}. This is due to the fact that it has the interesting property that the BDD size varies from linear to exponential dependent on the chosen variable ordering. 

\item[\boldmath\( f_3 = x_1x_2x_3 + x_4x_5x_6 + \cdots + x_{n-2}x_{n-1}x_n \):] A slight extension is to move from two to three variables per monomial. This function is quite similar to $f_2$ but now requires to group three variables in the base case and the induction step. Also here, LLMs work well (see {\bf Q3}). 

\item[Totally symmetric functions:] For totally symmetric functions (see {\bf Q4}) it is known that BDDs only require a quadratic number of nodes, i.e.~$O(n^2)$ \cite{TC1986}. The BDDs only grow by one node per level in the worst case, i.e.~one node in the first level, two in the second, etc. For many symmetric functions, like AND and OR, the final number is much smaller. In this case, the LLM correctly gives the core idea.
\end{description}
Of course, in all cases a formal proof that should be tool assisted is required. But this can easily be linked to the graph-based techniques from \cite{MEMOCODE2023,PhilTrans2025}.

To show that some of the discussions require a deeper understanding, we finally consider the multiplier function, where it can be formally proven that the BDD size is exponential independent of the chosen variable ordering \cite{TC1991}. Since the Q\&A is longer in this case, only the core parts are reported: 
\begin{quote}
{\bf Q:} {\em Can you also proof by induction that BDDs for the multiplication functions are polynomially bounded?}
\end{quote}
This generates the answer: 
\begin{quote}
{\bf A:} This is a great and subtle question. The short answer is:

No — we cannot, in general, prove by induction that BDDs for binary multiplication functions are polynomially bounded in size. In fact, in the worst case, Binary Decision Diagrams (BDDs) for multiplication functions can have exponential size.

Let’s examine this carefully.

...

Summary

    Multiplication functions (e.g., A × B) are not guaranteed to have polynomial-size BDDs.

    There is no general inductive proof that BDDs for multiplication are polynomially bounded.

    In fact, under most variable orderings, BDDs for multiplication require exponential space.
\end{quote}
{\em Most variables} is not correct, since it can be proven that it holds for all. A next question/prompt is send to clarify this.
\begin{quote}
{\bf Q:} {\em But many cases does not mean all, right? can you give me a case where you can avoid the exponential blow-up.}
\end{quote}
As a result, you get:
\begin{quote}
{\bf A:} ... Case: Careful variable ordering

For full multiplication A × B (both n bits), variable ordering matters greatly.

A poor ordering like:

$a_0$, $b_0$, $a_1$, $b_1$, $a_2$, $b_2$, …

leads to exponential size.

But a better ordering is:

$a_0$, $a_1$, …, $a_{n-1}$, $b_0$, $b_1$, …, $b_{n-1}$

or the reverse:

$b_0$, $b_1$, …, $b_{n-1}$, $a_0$, $a_1$, …, $a_{n-1}$

This allows the BDD package to exploit sharing and avoid recomputation of intermediate partial products.

Empirically (see Bryant 1986 and follow-ups), this ordering can significantly reduce BDD size — sometimes making it polynomial in practice for moderate n.
\end{quote}
From a complexity analysis point of view the last sentence is not correct, while the core message is clear: for a small number of variables the program might terminate within the given resources. 

\section{Conclusions and Future Work}

LLMs are a very powerful tool in the context of text generation. This also holds for generation of human-readable proofs. But the correctness has to be checked by some formal tools. This has already been observed in various applications leading to approaches for explanation models (see e.g.~\cite{NIPS2022,REW2022,NIPS2023,NAACP2025}).  

While the suggestions for proofs --- at least for induction --- are promising, a formal approach has to be applied. This can either be based on identifying patterns as suggested in \cite{MEMOCODE2023,PhilTrans2025} or on some explanation that is based on logical reasoning. Further interesting aspects to study are:
\begin{enumerate}
\item While the current induction proofs are for well-studied problems, however the same principles are also applicable to modern data structures (see e.g.~\cite{ATS2024_2}). Hence, LLMs can potentially benefit there as well.
\item The LLMs have used the prior works as reference while reasoning about the proofs in some cases. However, with the availability of {\em Retrieval Augmented Generation} (RAG), ongoing tailored proofs can be fed to the LLMs to build complex proofs that are human-readable.
\item Currently, even with simple prompts the reasoning capabilities of LLMs have shown potential. Using techniques like chain of thought reasoning and feedback from the formal tool can further aid the LLMs in proof generation.
\end{enumerate}
For PFV a linking of formal tools with the LLM is a promising direction for future work. 
\begin{credits}
\subsubsection{\ackname} This work was supported by the German Research Foundation (DFG) within
the Reinhart Koselleck Project PolyVer (DR 287/36-1 - project number 431649366) and the Research Training Group GRK 2972 CAUSE (project number 513623283). 
\end{credits}
%
%
%
%

\section{Appendix}\label{se:appendix}

\subsection*{Q1: How can I perform an induction proof for binary decision diagrams?}

\textbf{A:} To perform an induction proof involving Binary Decision Diagrams (BDDs), you typically use mathematical induction on a structural property (e.g., number of variables or size of the BDD). Here's a generic outline:

\begin{enumerate}[label=\textbf{Step \arabic*:}]
    \item \textbf{Base Case:} Prove the property holds for a trivial BDD, such as one with 0 or 1 variable.
    
    \item \textbf{Inductive Hypothesis:} Assume the property holds for all BDDs with up to \( n \) variables.
    
    \item \textbf{Inductive Step:} Prove it holds for \( n+1 \) variables. Use the fact that BDDs for \( n+1 \) variables can be split at the root into two BDDs with at most \( n \) variables each (Then/Else branches). Apply the inductive hypothesis.
\end{enumerate}

You might use this to prove correctness, size bounds, or structural properties of BDDs.

\subsection*{Q2: How does this work for the function \( f = x_1x_2 + x_3x_4 + \cdots + x_{2n-1}x_{2n} \)?}

\textbf{A:} Consider \( f_n = \sum_{i=1}^{n} x_{2i-1}x_{2i} \). Each term is a 2-variable AND. With a fixed variable ordering \( x_1 < x_2 < \dots < x_{2n} \), the BDD can be constructed incrementally.

\subsection*{Inductive Structure:}
\begin{itemize}
    \item Base Case (\( n = 1 \)): \( f_1 = x_1 x_2 \), BDD has 3 internal nodes.
    \item Inductive Hypothesis: BDD for \( f_k \) has size \( \leq 3k + 1 \).
    \item Inductive Step: Add \( x_{2k+1}x_{2k+2} \). New BDD adds 3 nodes. Total becomes \( 3(k+1) + 1 \).
\end{itemize}

Thus, the size is linear in \( n \).

\subsection*{Q3: How does it work for \( f = x_1x_2x_3 + x_4x_5x_6 + \cdots + x_{n-2}x_{n-1}x_n \)?}

\textbf{A:} Let \( n = 3k \), and define \( f_k = \sum_{i=1}^{k} x_{3i-2} x_{3i-1} x_{3i} \).

\subsection*{Inductive Strategy:}
\begin{itemize}
    \item Base Case: \( f_1 = x_1x_2x_3 \), BDD has 3 internal nodes.
    \item Inductive Step: Add new term \( x_{3k+1}x_{3k+2}x_{3k+3} \), which adds 4 nodes (including the OR node).
\end{itemize}

\subsection*{Conclusion:}
The BDD size grows linearly in \( k \). So for \( n = 3k \), the BDD size is \( O(n) \).

\subsection*{Q4: Can you also prove by induction that BDDs for totally symmetric functions are polynomially bounded?}

\textbf{A:} Yes. Let \( f_n \) be a totally symmetric Boolean function: it depends only on the Hamming weight of its input. We'll show the size of its BDD is \( O(n^2) \).

\subsection*{Inductive Argument:}
\begin{itemize}
    \item Base Case (\( n = 1 \)): BDD size is constant.
    \item Inductive Hypothesis: Assume \( f_n \) has a BDD with size \( \leq Cn^2 \).
    \item Inductive Step: Construct BDD for \( f_{n+1} \). At each level \( i \), track how many ones have been seen (\( 0 \leq k \leq i \)). Each level has \( i+1 \) nodes.
\end{itemize}

\subsection*{Node Count:}
\[
\sum_{i=0}^{n} (i+1) = \frac{(n+1)(n+2)}{2} = O(n^2)
\]

%
%

\end{document}